\documentclass[aps,prl,twocolumn,letterpaper,nobibnotes,superscriptaddress,longbibliography]{revtex4-1}
\usepackage{amssymb}
\usepackage{amsmath}
\usepackage{graphicx}
\usepackage{extarrows}
\usepackage{url}
\usepackage{varioref}
\usepackage{hyperref}
\usepackage{cleveref}
\usepackage{lineno}
\usepackage{color}
\usepackage{ulem}

\usepackage{float}
\usepackage{verbatim}

\begin{document}

\title{Observation of many-body quantum phase transitions beyond the Kibble-Zurek mechanism}

\author{Qi Huang}
\affiliation{School of Electronics Engineering and Computer Science, Peking University, Beijing 100871, China}
\author{Ruixiao Yao}
\thanks{Current address: Department of Physics and Research Laboratory of Electronics, Massachusetts Institute of Technology, Cambridge, MA, 02139, USA}
\affiliation{Department of Physics and State Key Laboratory of Low Dimensional Quantum Physics, Tsinghua University, Beijing, 100084, China}
\author{Libo Liang}
\affiliation{School of Electronics Engineering and Computer Science, Peking University, Beijing 100871, China}
\author{Shuai Wang}
\affiliation{Department of Physics and State Key Laboratory of Low Dimensional Quantum Physics, Tsinghua University, Beijing, 100084, China}
\author{Qinpei Zheng}
\affiliation{School of Electronics Engineering and Computer Science, Peking University, Beijing 100871, China}
\author{Dingping Li}
\affiliation{School of Physics, Peking University, Beijing 100871, China}
\author{Wei Xiong}
\affiliation{School of Electronics Engineering and Computer Science, Peking University, Beijing 100871, China}
\author{Xiaoji Zhou}
\affiliation{School of Electronics Engineering and Computer Science, Peking University, Beijing 100871, China}
\author{Wenlan Chen}
\email{cwlaser@ultracold.cn}
\affiliation{Department of Physics and State Key Laboratory of Low Dimensional Quantum Physics, Tsinghua University, Beijing, 100084, China}
\affiliation{Frontier Science Center for Quantum Information, Beijing, 100084, China}
\author{Xuzong Chen}
\email{xuzongchen@pku.edu.cn}
\affiliation{School of Electronics Engineering and Computer Science, Peking University, Beijing 100871, China}
\author{Jiazhong Hu}
\email{hujiazhong01@ultracold.cn}
\affiliation{Department of Physics and State Key Laboratory of Low Dimensional Quantum Physics, Tsinghua University, Beijing, 100084, China}
\affiliation{Frontier Science Center for Quantum Information, Beijing, 100084, China}

\begin{abstract}
Quantum critical behavior of many-body phase transitions is one of the most fascinating yet challenging questions in quantum physics. 
Here, we improved the band-mapping method to investigate the quantum phase transition from superfluid to Mott insulators, and we observed the critical behaviors of quantum phase transitions in both dynamical steady-state-relaxation region and phase-oscillation region.
Based on various observables, two different values for the same quantum critical parameter are observed.
This result is beyond a universal-scaling-law description of quantum phase transitions known as the Kibble-Zurek mechanism, and suggests that multiple quantum critical mechanisms are competing in many-body quantum phase transition experiments in inhomogeneous systems. 
\end{abstract}

\maketitle

Non-equilibrium quantum physics is one of the most challenging topics in quantum physics, related with many different areas such as quantum matters \cite{QTC, periodical}, many-body correlations \cite{MBL,MBL_huse}, and quantum simulations \cite{Simulation_1,Simulation_2, Turbulence_hydro,JoergKBMZ1,JoergKBMZ2}. 
One central question among the non-equilibrium quantum physics is how to understand quantum critical behaviors and dynamics of the quantum phase transitions (QPT). 
The Kibble-Zurek mechanism (KZM) \cite{KB, Zurek} originating from thermodynamics \cite{Ko_2019,Navon167,Formation,Ulm2013,Pyka2013}, describes dynamics of QPT with symmetry breaking \cite{phi4proposal,CChinKBMZ, LukinKBMZ, ChapmanKZ, DLMspin1,PhysRevLett.126.185302}, where the order parameter can be well-defined and the quantum critical behaviors have a universal scaling-law dependence on the external ramping speed.
As for the development of QPT, the energy gap {as well as} the symmetry breaking becomes one of the critical conditions of phase transitions. Thus, it becomes interesting to investigate many-body quantum phase transitions entering symmetry-conserved phases with open energy gaps, which is on the opposite side of the conventional KZM.

The superfluid (SF) to {Mott} insulators (MI) phase transition is one of the most important QPT in many-body physics.
Seminal experiments \cite{PRLMISF, PNASSchneider} explored the KZM based on transitions from symmetry-conserved MI to symmetry-broken SF with gap closing, where the measured quantum critical parameter are not consistent with the theoretical predictions \cite{BHMFisher,sachdev_2011}. 
Here, we investigate the many-body QPT, where the system transits from the SF phase (gapless symmetry-broken) to the MI phase (gapped symmetry-conserved) with an improved band-mapping method.
In such gap-opening QPT, the quantum criticality is described by the energy gap $\Delta$ versus the distance $g$ to quantum critical point {in phase diagrams} with a scaling law $\Delta\propto |g|^{\nu z}$ where $\nu z$ is a quantum critical parameter. The value of $\nu z$ decides the spatial-temporal universal dynamics in KZM \cite{CChinKBMZ,2012EDemler,NJPKuno_KZ,PRAKuno_KZ,KBMZorigin2005,PRL_miguel,NJP_miguel}.
Based on various observables, we extract out two different values for the same critical parameter $\nu z$.
Both values were predicted theoretically \cite{BHMFisher,sachdev_2011}, but the coexistence of two different values for the same parameter violates the universality of KZM.
Besides, we observe dynamical steady-phase relaxations within the KZM frame transiting into non-steady-phase oscillations which is beyond the KZM frame. 
In fact, the open gap in the many-body systems allows different values of critical parameters to exist in the same QPT, protects {phase-oscillations} and thus boosts the many-body QPT beyond the conventional symmetry-breaking KZM.

\begin{figure*}[htbp]
	\centering
	\includegraphics[width=16cm]{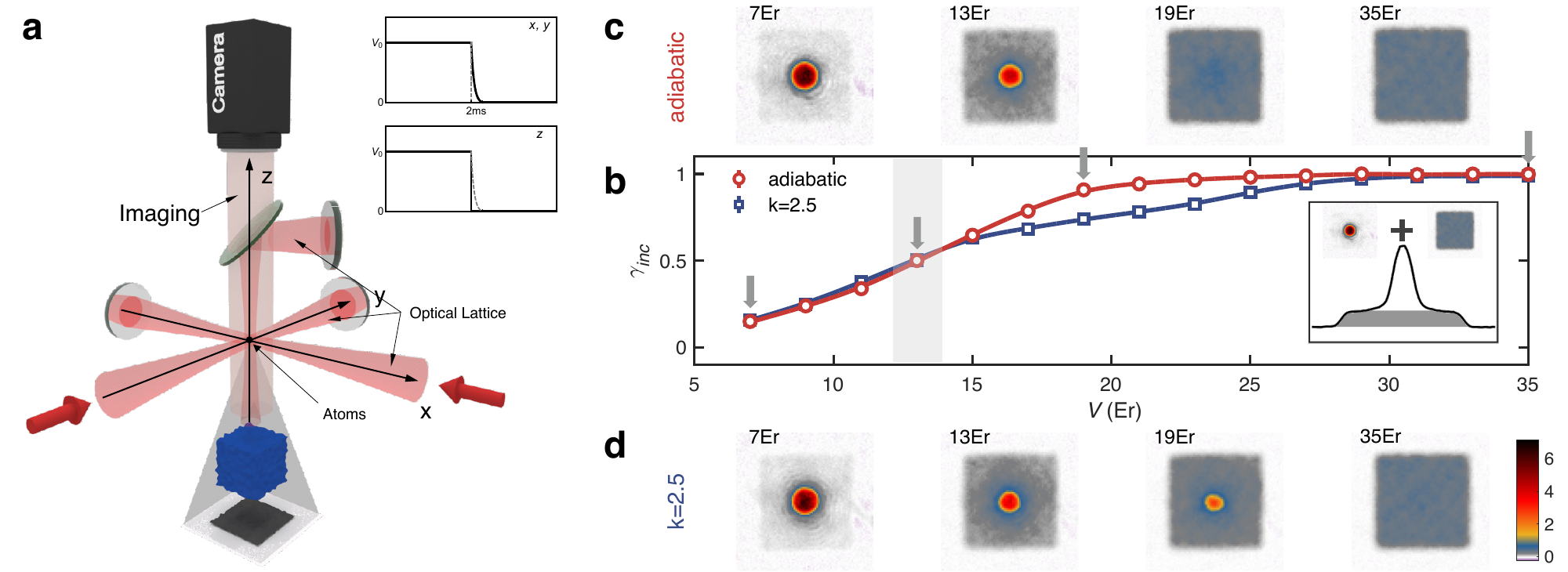}
	\caption{\textbf{Experimental setup and improved band-mapping method.} \textbf{a}, Three orthogonal standing waves form 3D optical lattices for $^{87}$Rb, and absorption images are taken along $z$ direction. The inset: the $z$-lattice is turned off instantaneously while the $x$- and $y$-lattices are ramped down in 2~ms for the band mapping.  
	\textbf{b}, The incoherent fraction $\gamma_{inc}$ versus the trap depth $V$ for the SF-MI phase transitions. {The red circles correspond to the measurement with adiabatic ramping, while the blue squares correspond to the measurement with linear ramping at rates $k=2.5E_r$/ms.}
The shadow area locates quantum phase transition at $V_c = 13E_r$ for $\bar{n} = 1$'s lobe.
Grey arrows indicate data in the panel c and d. The inset shows the decomposition of quasi-momentum profiles and analysis of $\gamma_{inc}$. 
	{\textbf{c} and \textbf{d}} show the band mapping profiles performed at trap depth $V=$ 7, 13, 19, and 35$E_r$.
Error bars (one standard deviation) are smaller than the marker size.}	
	\label{F1}
\end{figure*}

Our experiment is performed in a three-dimensional optical lattice formed by three standing waves perpendicular to each other at wavelength $\lambda=1064$~nm (Fig.~\ref{F1}\textbf{a}), and the magnetic field is applied along $z$ axis. We prepare Bose-Einstein condensates of $^{87}$Rb atoms in $|F = 1,m_F = -1\rangle $ and then load them into 3D homogeneous lattices with trap depth $V=5E_r$ and atom number $N=1.1(2) \times 10^5$ \cite{zhou2019dimension,Tianwei,TianCooling}. Here, $V$ is the trap depth generated by one lattice beam and $E_r=h\times 2$~kHz is the recoil energy of lattice beams.
Due to the Gaussian shape of lattice beams, the lattices are printed by an external harmonic trap with homogeneous radial vibrational frequencies $\sim 2\pi \times {20(1)}$~Hz. The system is described by a Bose-Hubbard model in an external harmonic trap \cite{BHMFisher,Markus03}:
\begin{eqnarray}
H &=&-J \sum_{\langle i,j \rangle}(\hat{a}_i^{\dagger} \hat{a}_j + h.c.) +\frac{1}{2}U \sum_{i}\hat{n}_i(\hat{n}_i-1) \nonumber \\
& &+ \sum_{i}(\frac{1}{2}m\omega^2_0 r_i^2 - \mu)\hat{n}_i,
\label{BHM}
\end{eqnarray}
where $a_i$ (or $a^\dagger_i$) is the annihilation (or creation) operator of a boson at the lattice site $i$, $J$ is the tunneling coefficient, $U$ is the on-site interaction, $\mu$ is the chemical potential, and ${1\over 2}m\omega^2_0 r^2_i$ describes the external harmonic trap. {Since the atomic size is much smaller than the waist of lattice beams, the inhomogeneity of $J$ and $U$ are negligible in this model (see Supplementary Information Part-I (SI-I)).}

Comparing with the previous band-mapping method \cite{Esslinger_band_mapping} requiring adiabatically turning off three lattice beams, here we instantaneously turn off the $z$-lattice that is along the imaging direction (Fig.~\ref{F1}\textbf{a}). 
The atomic gas expands along $z$ quickly, releases the interaction energy, and the momentum distribution along $z$ does not contribute to the band-mapping image. 
It helps to avoid the fast relaxations in the $x$-$y$ plane in the subsequent expansion. 
At the same time when the $z$-lattice is turned off, we ramp down the $x$- and $y$-lattices in 2~ms, which is slow enough to adiabatically convert the quasi-momentum into the real momentum. 
Then the time-of-flight is applied to measure the momentum distribution in the $x$-$y$ plane.
Using this technique, we are able to obtain better quasi-momentum distributions of atoms in the interacting system (Fig.~\ref{F1}\textbf{b} and SI-II), where the incoherent atoms in the first Brillouin zone are exactly mapped to a square shape and displays a flat plateau, while the phase coherent superfluid component corresponds to a narrow and sharp peak at the zero-momentum point. 
This help us to quantify the coherence across the whole system (Fig.~\ref{F1}\textbf{b}) {and to observe the dynamical response at different ramping rates (Fig~\ref{F1}\textbf{c} and \textbf{d})}. 
We define the incoherent fraction $\gamma_{inc}$, as the ratio of the integrated optical depth of the flat plateau divided by the total integrated optical depth (Fig.~\ref{F1}\textbf{b} and SI-III) based on the band-mapping images.

\begin{figure*}[htbp]
	\centering
	\includegraphics[width=16cm]{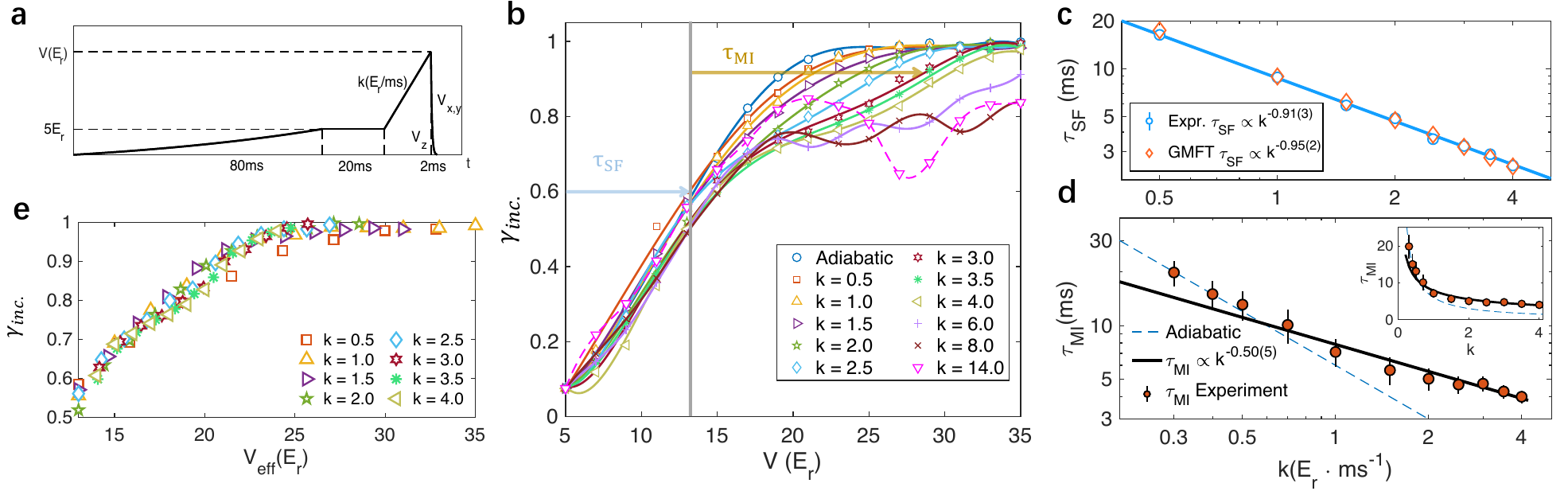}
	\caption{\textbf{Dynamical response of SF-MI phase transitions.} \textbf{a}, The time sequence of the trap depth ramping. The first stage of 80~ms and the second stage of 20~ms prepare superfluid samples from the condensates. The third stage is the linear ramp with a ramping rate $k$ when the atoms experience phase transitions. The final stage is the band mapping to distinguish the coherent component from the incoherent one. \textbf{b}, The incoherent fraction $\gamma_{inc}$ versus trap depth $V$ for different $k$. The scattered markers are the experimental data and the solid lines are the polynomial fits (see SI-{V}). A furcation appears at critical point $V_c = 13E_r$. And $\gamma_{inc}$ approaches 1 for different ramping rate $k$ for small $k$. When $k$ gets larger, $\gamma_{inc}$ starts to oscillate with retard thermalization. {The pink triangles denote the dynamics at $k = 14E_r$/ms, which oscillates drastically and fail to relax}. Here we label the definition of SF response time $\tau_{SF}$ (time to reach $\gamma_{inc}=0.6$) and the MI dynamical relaxation time $\tau_{MI}$ (time between $\gamma_{inc}=0.6$ and $\gamma_{inc}=0.9$) on the plot. 
	\textbf{c}, $\tau_{SF}$ versus $k$. The blue circles are the experimental data and the blue solid line is the fit based on power laws with $\tau_{SF}\sim k^{-0.91(3)}$. The red diamonds are the GMFT simulation 
with a fit $\tau_{SF}\propto k^{-0.95(2)}$. 
\textbf{d}, $\tau_{MI}$ versus $k$. The red filled circles are the extracted data. {The blue dashed line is $\tau_{MI}\sim k^{-1}$ for the adiabatic response (see SI-VI).} The black solid line is the fit with $\tau_{MI}\sim k^{ {-0.50(5)}}$ {for $k\geq 0.7E_r$/ms}, which indicates $\nu z={1.0(2)}$.
\textbf{e}, The universal response of $\gamma_{inc}$ versus the rescaled trap depth $V_{eff}$, where 
{$V_{eff}= (V-V_c)k^{-(1-{0.50})} + V_c$} and $k$ is in the unit of $E_r$/ms. Error bars (see SI-V) correspond to one standard deviation.}
	\label{F2}
\end{figure*}

To study dynamics of QPT, we first prepare the superfluid at $V_0 = 5E_r$ and hold it for 20~ms. Then, we ramp up the trap depth $V$ linearly with a ramping rate $k$ (Fig.~\ref{F2}\textbf{a}). For each $k$, we perform the band mapping at different $V$ and measure the corresponding incoherent fraction $\gamma_{inc}$ (Fig.~\ref{F2}\textbf{b}). According to Ref. \cite{Markus03,QMC}, the MI starts to appear at $V_c=13 E_r$ for $^{87}$Rb in 3D optical lattices.
We find two different response regions before and after ramping through $V_c$.
In the SF region, the system responses to the ramping rapidly due to its gapless nature.
We measure the response time $\tau_{SF}$ by defining the time to reach $\gamma_{inc}=0.6$ and obtain a scaling-law dependence $\tau_{SF}\propto k^{-0.91(3)}$, slightly delayed from the ideal instant response $ k^{-1}$. 
We performed numerical calculations by the Gutzwiller mean field theory (GMFT) \cite{Gutzwiller03, Kuba2005Gutzwiller}, finding it consistent with the experimental results (Fig.~\ref{F2}\textbf{c}). 
However, once the trap depth is above $V_c$, $\gamma_{inc}$ starts to furcate  
and show retarded or oscillating responses depending on $k$.

For a slow ramp ($ k\le 4 E_r/$ms), the response in the MI region is trying to approach the steady state with a dynamical relaxation time $\tau_{MI}$. In Fig.~2\textbf{d}, we extract $\tau_{MI}$ based on the time duration to reach 
$\gamma_{inc} = 0.9$ from the time at trap depth $13 E_r$ \cite{CChinKBMZ}. 
It shows a scaling-law dependence {of non-adiabatic response $\tau_{MI}\propto k^{-0.50(5)}$ for $k\geq 0.7E_r$/ms.} 
In SI-{VI}, we verify that the exponent near $-0.5$ is not sensitive to the end point $\gamma_{inc}$ we choose.
The KZM predicts the dependence of the {freeze-out time} $\tau$ on the ramping rate $k$ to be \cite{CChinKBMZ, 2012EDemler,NJPKuno_KZ,PRAKuno_KZ}
\begin{equation}
\tau \propto k^{-\frac{\nu z}{1+\nu z}},
\end{equation} 
{where $\tau$ characterizes the time of relaxation.} Based on this form, we obtain the critical parameter $\nu z={1.0(2)}$ by measuring the MI {dynamical} relaxation time, and infer $\nu=1/2$ \& $z=2$ corresponding to the off-tip critical parameters with linear gap opening in the SF-MI phase diagram \cite{BHMFisher}. 
To show the quantum criticality, we rescale the horizontal axis of Fig.~\ref{F2}\textbf{b} at $V_c$ to be $V_{eff}= (V-V_c)k^{-{(1-0.50)}} + V_c$. The different sets of $\gamma_{inc}$ data fall into one {universal curve (Fig.~\ref{F2}\textbf{e})} in the MI region. This shows the universality of the KZM if we consider only $\tau_{MI}$ under a slow ramp $k$. 
However, GMFT cannot describe this relaxations (see SI-VII) due to strong interactions of deep MI \cite{QMC}.

\begin{figure}[htbp]
	\centering
	\includegraphics[width=8 cm]{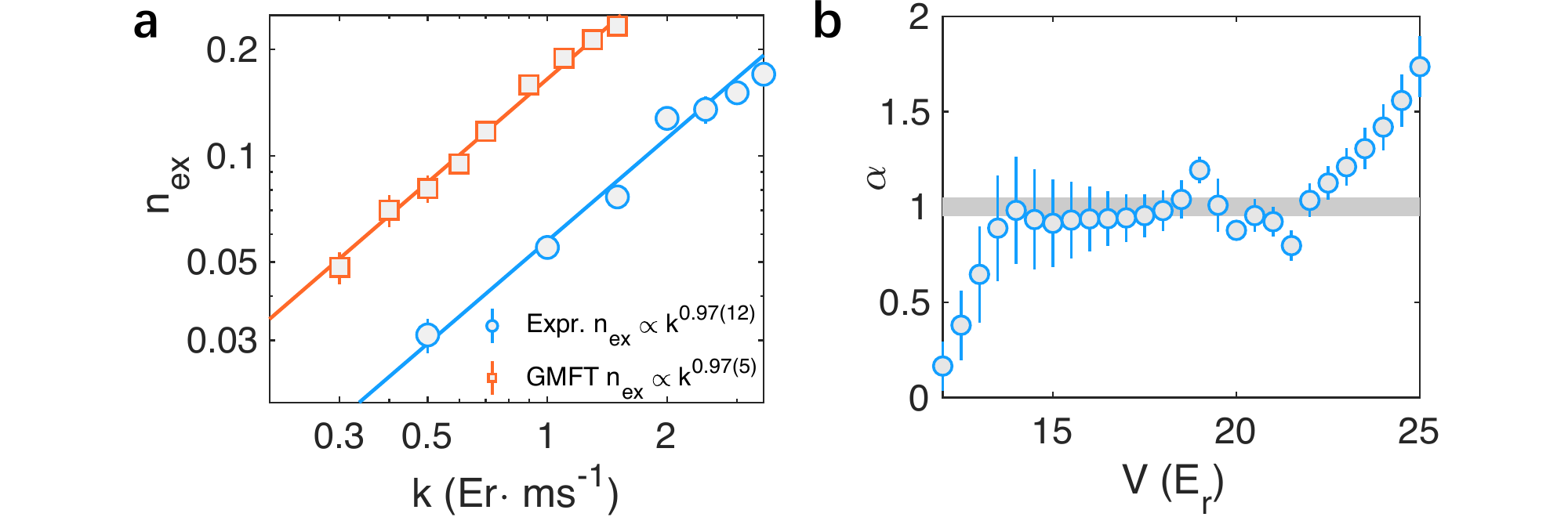}
	\caption{\textbf{Excitation fraction $n_{ex}$ in the MI region.}	
	\textbf{a,} $n_{ex}(k)$ measured at $V \sim 18E_r$ (SI-{VI}). 
	Blue circles are experimental results with the blue line fit of $n_{ex}\sim k^\alpha$ at $\alpha=0.97(12)$. Orange squares are GMFT simulation results with the orange line fit of $n_{ex}\sim k^\alpha$ at $\alpha=0.97(5)$.
	\textbf{b}, The fitted scaling coefficient $\alpha$ is robust against the chosen trap depth $V$, as long as $V > 13E_r$ (goes through phase transition) and $V < {19}E_r$ (far from deep MI region). Error bars correspond to one standard deviation.}
	\label{F3}
\end{figure}

\begin{figure*}[htbp]
	\centering
	\includegraphics[width=16cm]{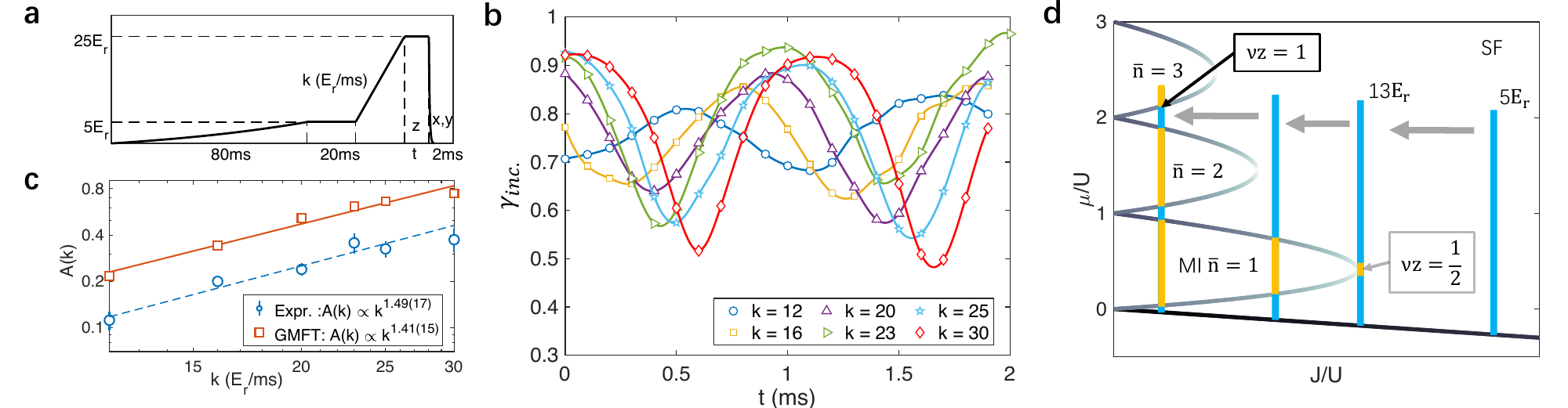}
	\caption{\textbf{Phase-oscillations under a fast ramp.}	
	\textbf{a}, Time sequences for oscillation measurements. The atoms are hold at $V=25E_r$ for a varying time $t$.
	\textbf{b}, $\gamma_{inc}$ oscillates versus time $t$ for different $k$. The solid lines are cubic splines fit. \textbf{c}, The fitted oscillation amplitude $A$ versus $k$. Blue circles are experimental data and the red squares are GMFT simulations. Two dashed lines are the fits based on power laws. For experimental data we obtain $A(k) \propto k^{1.49(17)}$, and the GMFT gives $A(k) \propto k^{1.41(15)} $. \textbf{d}, An illustration of the dynamics in the SF-MI phase diagram. The blue lines represents our actual system and the on- or off-tip locations are labeled by arrows with $\nu z=1/2$ or 1. The on-tip location has square-root gap opening and the off-tip one has linear gap opening.
	} 
	\label{F4}
\end{figure*}

Besides a scaled relaxation time, the KZM also predicts universally scaled defect density introduced by crossing the phase transition. Here we use
the excitation fraction $n_{ex}$ to characterize the defects, where it describes how many atoms are excited comparing to 
ground-state Mott insulators. $n_{ex}$
follows the form of \cite{KBMZorigin2005, 2012EDemler,PRL_miguel,NJP_miguel}:
\begin{equation}
n_{ex}\propto k^{{d\nu \over1+\nu z}}. \label{nex}
\end{equation}
Here $d=3$ is the dimension of 3D optical lattices. We analyze the excitation fraction as the components of MI fractions deviated from its adiabatic value \cite{PRLMISF}: $n_{ex}(V,k)=\gamma_{inc}(V, adia)-\gamma_{inc}(V,k)$. For $V>V_c$, this definition quantifies the additional particle-hole pairs on top of the MI ground state with quantum fluctuations, which retains additional phase coherence due to the non-equilibrium dynamics {(see SI-VII for more analysis)}.
In Fig.~\ref{F3}\textbf{a}, we plot $n_{ex}(V,k)$
and obtain a scaling-law $n_{ex}\propto k^{0.97(12)}$. 
The value of $n_{ex}(V,k)$ depends on the chosen trap depth $V$, but the value of fitted-exponent $\alpha$ is robust against different $V$ (Fig.~\ref{F3}\textbf{b}). 
The fitting on Eq. \ref{nex} shows $\nu z=0.48(9)$, 
where we infer $\nu =1/2$ \& $z=1$, corresponding to the on-tip critical {parameters} with square-root gap opening in the SF-MI phase diagram \cite{BHMFisher}. 
To better understand this result, we apply GMFT in {shallow MI region}
and find the defect density due to the non-equilibrium dynamics has a scaling factor of 0.97(5), consistent with our observations (see SI-{VII}).

For a fast ramp ($k\ge 6E_r/$ms), the system response enters a non-steady-state region where $\gamma_{inc}$ oscillates with time.
This provides a smooth connection from the adiabatically ramping to the fast ramping \cite{Revival}. In this region,
the multiple-atom occupancies in the SF are frozen 
rapidly without relaxations. However, the phase coherence between single- and multiple-occupancies does not disappear immediately and oscillates at the frequency of $U/\hbar$.
To visualize the oscillation, we hold the lattice for a time interval $t$ after the ramping, where a larger $k$ leads to a larger oscillation amplitude $A$ (Fig.~\ref{F4}\textbf{a} and \textbf{b}).
The fit shows {$A\propto k^{1.49(17)}$} (Fig.~\ref{F4}\textbf{c}). In GMFT we see similar trends, where {$A\propto k^{1.41(15)}$} is obtained. 
Since we stop ramping at deep MI region at $V = 25E_r$, the oscillation amplitude $A$ characterizes multiple occupancies as the excitation 
for such non-steady-states ramping across phase transition. 
Since this lack of thermalization and subsequent oscillation is the direct consequence of the gap-opening process, the scaling dependence $A\sim k^{3/2}$ provides a new relation in the non-steady-state region which is linked to the quantum phase transition but was not explored by conventional KZM.

Based on different observables, we get two different values for the same {quantum critical parameter} $\nu z$, namely $\nu z=1$ for the MI {dynamical} relaxation time $\tau_{MI}$ and $\nu z=1/2$ for the excitation fraction $n_{ex}$.
According to conventional KZM, 
the universality of the quantum criticality 
predicts the same value of $\nu z$ for different observables in the same QPT.
Thus, our measurement results contradict the conventional KZM. 
According to previous studies \cite{BHMFisher,sachdev_2011}, $\nu z$ is either 1 or 1/2 depending on whether the phase transition point is off- or on-tip and how the gaps open in the SF-MI phase diagram (Fig.~\ref{F4}\textbf{d}). 
Due to the harmonic trap and fixed atom number in our experiment, we probe the phase transition with an unfixed chemical potential, corresponding to a line, not a point in the phase diagram. {The center filling of the final stage of MI is approximately $\bar{n} \sim 3$.}
In the scenario of a slow ramp, the MI appear {and coexist with SF while} the trap depth is not deep enough.
Because the local gap opens as $\Delta\propto |g|^{1/2}$, the early-formed MI defects appear in {the region of MI} under the quantum criticality of $\nu z=1/2$.
Later when the trap depth increases, 
the total system enters the region of off-tip phase transitions where the gap $\Delta \propto |g|$ opens linearly with $\nu z=1$. However, the previously-generated defects are still protected by the energy gaps under $U(1)$ symmetries 
with $\nu z=1/2$.
This is how both $\nu z$ appear in the same QPT experiment.
{If the QPT is performed in the opposite direction from gapped phases to gapless phases as in the previous experiments \cite{PRLMISF, PNASSchneider}, the gapless excitations due to broken symmetries will smear out early-formed defects with later dynamics and makes the different values of quantum critical parameter indistinguishable.} 
For a fast ramp, the system is immediately frozen into deep Mott insulators with failed relaxations due to large energy gaps, where the actual dynamics {versus the excitation fractions or the correlation length} is still an unexplored regime for non-equilibrium physics and waits for further investigations.

In conclusion, we observe the many-body quantum phase transition from gapless symmetry-broken phases to gapped symmetry-conserved phases. 
As the ramping gets faster, the critical behaviors change from retarded relaxations to phase oscillations. Even within the steady-state-relaxation regime, two different values for one critical parameter $\nu z$ are observed.
The gap opening provides and protects different critical-mechanism competitions in the dynamics of many-body phase transitions. 
We believe this inspires further investigations in the phase transition of symmetry-conserved gapped systems and the non-equilibrium physics.

We thank Hui Zhai,  Joerg Schmiedmayer, Jakub Zakrzewski and Bogdan Damski for helpful discussions. The numerical simulations are performed on High-Performance-Computing Platform of Peking University. This work is supported by the National Natural Science Foundation of China (Grants No. 91736208, 11920101004, 61727819, 11934002, 11974202, 61975092) and the National Key Research and Development Program of China (Grant No. 2016YFA0301501).

\bibliography{references}

\widetext
\newpage

\begin{center}
\large\textbf{Supplemental Information for Observation of many-body quantum phase transitions beyond the Kibble-Zurek mechanism}
\end{center}
\setcounter{figure}{4}
  
 \tableofcontents 
  
\subsection*{I. The uncertainty of trap depth and the inhomogeneity of $J/U$}
The uncertainty of trap depth comes from two factors: (i) the calibration errors of trap depth; (ii) the fluctuations of the laser intensity. We use the method of multiple pulses Kapitza-Dirac diffraction along with BPNN algorithm \cite{Tianwei} to calibrate the trap depth, reaching a relative uncertainty of $0.6\%$. The laser intensity is controlled within fluctuations of $0.7\%$. Therefore, the overall relative uncertainty of the trap depth is below $1.3\%$.

{
In our system, the waist of lattice beams is 150(10) $\mu$m while the atoms occupy a region with radius of 13 $\mu$m. The intensity of lattice beams at the atomic wings decrease to $\exp\left(-2\times13^2/150^2\right)=98.5\%$. Thus, at $V=13E_r$, the tunneling rate $J$ increases by 4.2\% and the interaction strength $U$ decreases by 1.3\%, due to this 1.5\% change in the trap depth. The ratio of $J/U$ increases by 5.3\%. 
At $V=25E_r$, $J$ increases by 6.5\% and $U$ decreases by 1.3\%, due to this 1.5\% change in the trap depth. $J/U$ increases by 7.4\%. 
Comparing to the local chemical potential difference between wings and center, the differences of $J/U$ at each particular trap depth are negligible. Therefore, the inhomogeneous behaviors of Bose-Hubbard models are dominated by the inhomogeneity of the local chemical potentials.
}

\subsection*{II. Improved band mapping}
The conventional band-mapping method requires adiabatic ramping down of all lattice beams for a 3D optical lattice. The purpose of adiabatic ramping is to convert the lattice quasi momentum into the free-space real momentum. 
In the Mott insulators phase, each atom is tightly trapped within one site which corresponds to the superposition of all the Bloch function with possible momenta within one band (Wannier function). Therefore, {the band-mapping image} is expected to
be a uniform distribution over the first Brillouin zone. 

\begin{figure}[htbp]
	\centering
	\includegraphics[width=14 cm]{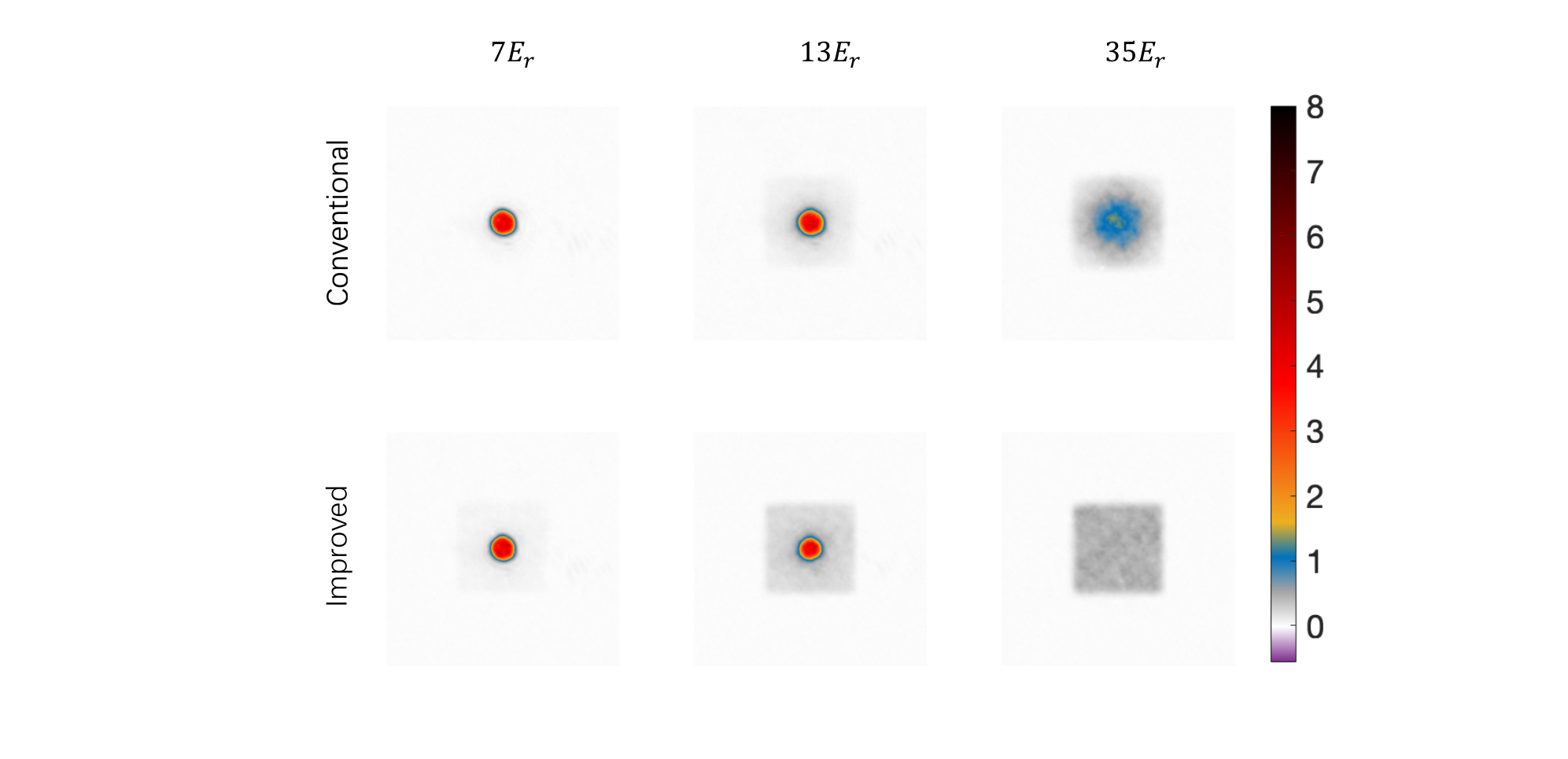}
	\caption{\textbf{Improvement of band mapping.} The conventional and improved {band mapping} measurement are plotted at $V = 7E_r$, $13E_r$ (critical {point}), and $35E_r$ (deep MI). In the SF region, two methods give similar results. {As} interaction increases, our method starts to shows a much-flatter plateau. When the system reaches the deep {MI region}, our method gives {uniform} momentum distribution in the first Brillouin zone. For comparison, the conventional method gives a central peak in the {center which} cannot be fully distinguished from the superfluid peak.}
	\label{SI_BM}
\end{figure}

However, due to the strong interaction in the Mott insulators region, the on-site interaction energy $U$ alters the final momentum distributions during the time-of-flight imaging. 
In our improved version of band-mapping, we turn off the $z$ lattice instantaneously and ramp down the $x$ and $y$ {lattices}. The interaction energy is released into the image integration direction without affecting the other two directions, and this {release leads to} a better conversion from the quasi-momentum in the first Brillouin zone into the real momentum in the free space. Here, we plot Fig.~\ref{SI_BM} to directly compare {results of both band-mapping} methods at different trap depth. Our method gives an exact square shape with a flat plateau for deep Mott insulators. It also gives a better resolution to distinguish {coherent and incoherent parts} in the shallow Mott insulator region.

\subsection*{III. Analysis of the incoherent fraction}
\begin{figure}[htbp]
	\centering
	\includegraphics[width=14 cm]{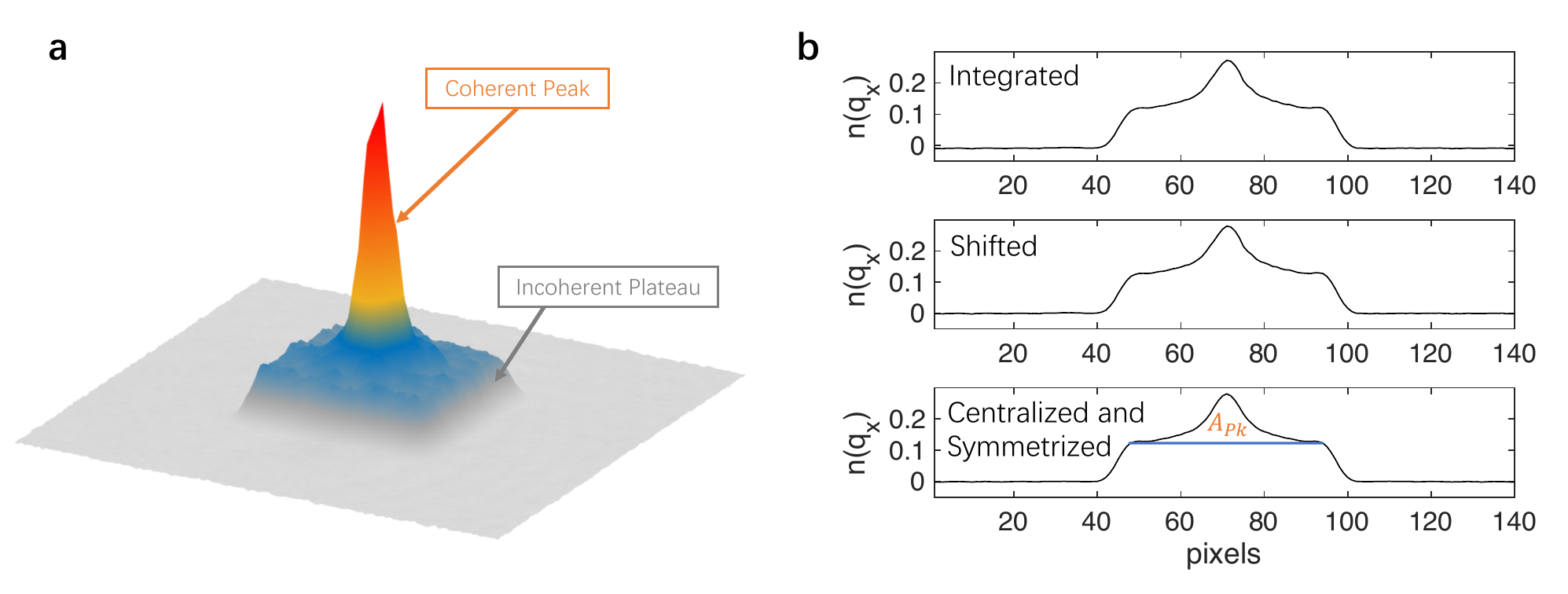}
	\caption{\textbf{How to {measure} $\gamma_{inc}$.} \textbf{a}, A band mapping profile consisting of two components, {the central peak contributed by coherent atoms and the plateau contributed by incoherent atoms.} \textbf{b}, {Distribution with} integration {along $y$ axis}, background removal, and centralization with symmetrization. {$A_{pk}$}  is defined in the lowest panel.}
	\label{process}
\end{figure}

The {incoherent fraction $\gamma_{inc}$} is defined as the fraction of the flat plateau in the quasi-momentum distribution. In the practical data processing, we {quantify the fraction of} the $q = 0$ {coherent peak} and subtract it from 1 to {get} {$\gamma_{inc}$}. We integrate the quasi-momentum distribution $n(q_x,q_y)$ (Fig.~\ref{process}\textbf{a}) along $y$ to obtain $n(q_x)$, as in the inset of Fig.~1\textbf{c}. 
To illustrate how we process the data, we plot Fig.~\ref{process}\textbf{b} to help us explaining each step. We exclude the background value, i.e. shift $n(|q_x| > \pi)$ to be zero. 
In the measurement of the band mapping, the central peak has a finite width due to {finite time of flight}. Because our method can give a flat plateau in the deep MI region,
we define the height of plateau to be $n(q_x \sim \pm \pi)$, and all deviations on top of this rectangle are regarded as the contributions of the {coherent component}.
To avoid possible in-alignment errors, we re-define quasi-momentum zero to be the center of mass of the $n(q_x)$, and the symmetrization has been performed as ${n_{q_x}} \rightarrow (n_{q_x} + n_{-q_x})/2$. We obtain the area of {coherent peak} $A_{pk}$ and total area $A_{tot.}$. The {incoherent fraction} is {defined} as:
\begin{equation}
\gamma_{inc} = 1- \frac{A_{pk}}{A_{tot.}}.
\end{equation}
Here {$\gamma_{inc}$} describes the fraction of incoherent atoms. The increase of {$\gamma_{inc}$} means the disappearance of phase coherence. 

{
If the temperature of atoms was much higher than the band width of the first Brillouin zone, the band mapping method would also show a flat plateau in the time-of-flight imaging. However, this is not the case in our experiment. We apply an improved cooling method \cite{TianCooling} to reach a low-enough initial temperature and this avoids the uniform thermal occupations in the first Brillouin zone. Therefore, the incoherent fraction $\gamma_{inc}$ is mostly contributed by the many-body phase transitions, not the thermal effects.
}

 {
 \subsection*{IV. Discussion of KZM from symmetry-broken phases to symmetry-conserved phases}
We adapted the results from Ref. \cite{KBMZorigin2005}, where the Kibble-Zurek mechanism should work in both directions for many-body phase transitions. The validity of KZ dynamics in SF-to-MI transitions is supported by the fact that the typical energy $\Delta$ scales in the same polynomial way $\Delta\propto|g|^{\nu z}$ near the quantum critical point for both sides. In superfluid, the typical energy $\Delta$ is the excitation energy for a Bogliubov mode with momentum $k\sim1/\xi$, where $\xi$ is the healing length. In Mott insulators, $\Delta$ is the energy gap to create a particle-hole pair. The scaling of KZ dynamics depends on the energy scaling relation \cite{2012EDemler} and behaves the same for both directions ramping between symmetry-broken phases (superfluid) and symmetry-conserved phases (Mott insulators). 
 }
 
 \subsection{V. Error bar estimations and fitting line in Fig.~2 and 4}
For each data point in Fig.~2 and 4 in the main text, we repeat the same measurements for $18$ times, and group the samples as $3\times 6$. We average each $3$ band mapping profiles and analyze {$\gamma_{inc}$}. The expected value and standard deviation of one specific data point are evaluated for such $6$ groups. 
Based on our band-mapping method, the one standard deviation error bar is usually under $1\%$ and smaller than the marker size.
 
In Fig.~2 and Fig.~4, the dynamics of various ramping rate $k$ {below} $4E_r$/ms are fitted with polynomial with 6th order, while for $k > 4E_r$/ms discrete data points are connected by spline interpolation. 
 For $\tau_{MI}$ in Fig.~2, we cut at $\gamma_{{inc}} \in [0.85,0.95]${, and} then average the {obtained} delay time $\tau_{MI}$. 
 One could also fit every independent cut among the interval, {and} then average the exponents obtained. 
 We have examined {both} method{s and get the same result of exponents within error bars}. 
 For $n_{ex}$ in Fig.~3, the vertical cut are set to be $V/E_r \in [18,20]${, and} then we perform the same data processing stated above. The amplitude $A(k)$ in Fig.~4 is {analyzed} by fitting $\gamma_{{inc}}(t)$ versus $t$ in first $1.2$~ms {with expression} $\gamma_{{inc}}(t) = A\cos(Bt + C) + D$, with error bar denoting fitting error.

\subsection*{VI. Determination for Kibble-Zurek exponents in Fig.~2 and 3}
\begin{figure}[htbp]
	\centering
	\includegraphics[width=12 cm]{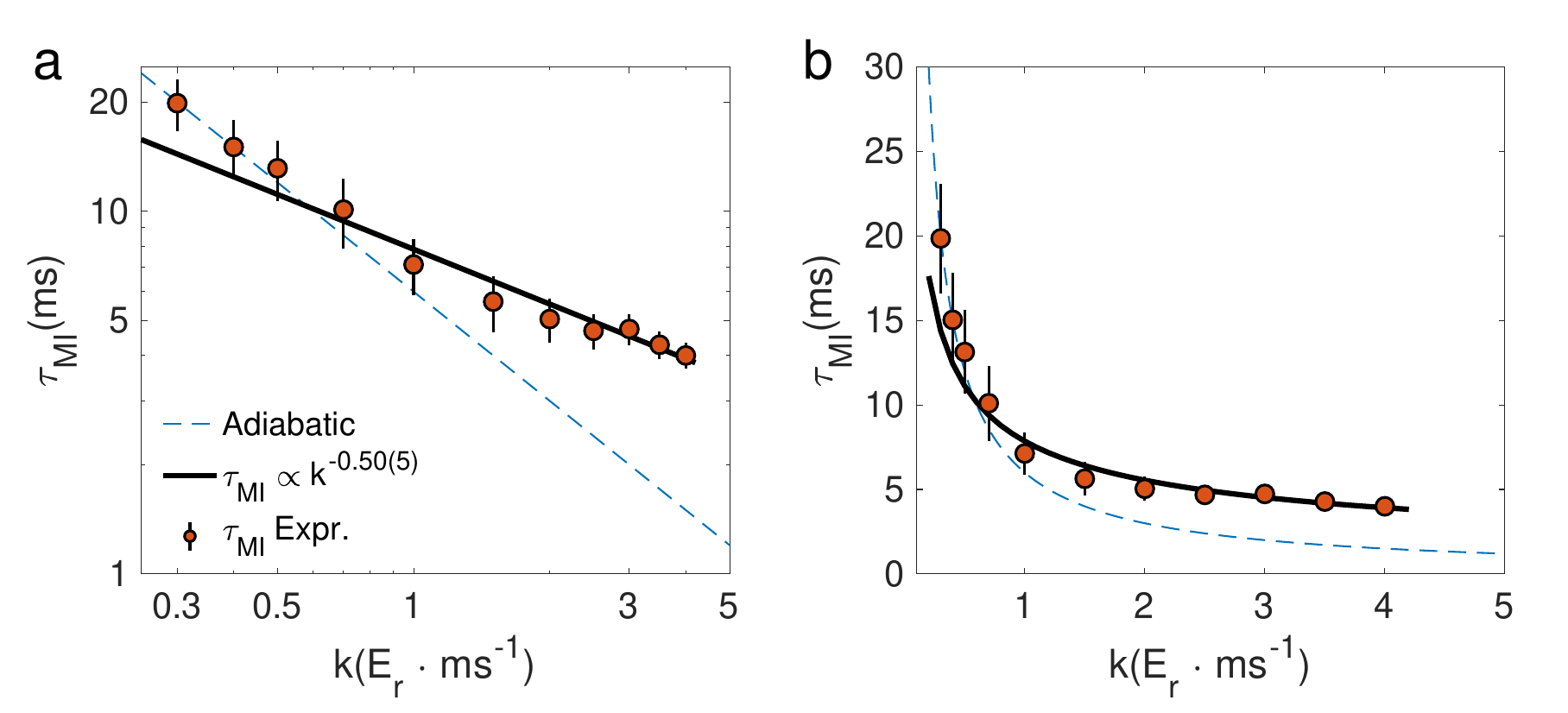}
	\caption{{\textbf{The adiabatic response and the delayed relaxation.} Here we show a larger version of Fig.~2d in the main text. The panel a is plotted in the logarithmic scale and the panel b is plotted in the linear scale. The red points are the data points. The blue dashed lines are the adiabatic response of $\tau_{MI}\sim k^{-1}$ calculated by the ratio of $\Delta V$ and $k$. It is consistent with the data points for $k\leq 0.7E_r$/ms. When $k$ becomes larger, the data deviates from the blue dashed line and shows a new trend. The solid black line is a fit of the data for $0.7E_r$/ms$\leq k\leq 4E_r$/ms, showing $\tau_{MI}\sim k^{-0.50(5)}$. The data show that there are two different regions for slow ramping of $k$, one is the adiabatic region and the other is the KZM region.}
	}
	\label{limi}
\end{figure}

{
In Fig.~2d of the main text, we plot $\tau_{MI}$ versus $k$. Here, we plot a larger version in Fig.~\ref{limi} to show different regions of the dynamical response. When the external ramping is very slow in the adiabatic region ($k\leq 0.7E_r$/ms), the measurement of $\gamma_{inc.}$ shows almost no difference from the adiabatic ramping and follows the ramping parameter instantaneously. Therefore, the response time $\tau_{MI}$ is the ratio of the trap depth changed $\Delta V$ divided by the ramping rate $k$. It leads to a dependence of $\tau_{MI}\sim k^{-1}$. In Fig.~2d and Fig.~\ref{limi}, the dashed blue line is the adiabatic response calculated by the ratio of $\Delta V$ and $k$, and it is consistent with the data of our system for $k<0.7E_r$/ms. }

{
When $k$ becomes larger ($0.7E_r$/ms$\leq k \leq 4E_r$/ms), the data points deviate from the adiabatic limit where $\gamma_{inc.}$ shows a larger response time $\tau_{MI}$. In this region, the system is described by the Kibble-Zurek mechanism that the system cannot follow the external ramping instantaneously and needs additional time to relax. When $k$ is further increased, we will start to see the oscillation of $\gamma_{inc}$. Therefore, we fit the response time of this region, deviating from the adiabatic limit but without oscillations, to capture the delayed relaxation behaviors. The black lines in Fig.~2d and Fig.~\ref{limi} are fitting lines of the data, being $\tau_{MI}\sim k^{-0.50(5)}$.
}

\begin{figure}[htbp]
	\centering
	\includegraphics[width=14 cm]{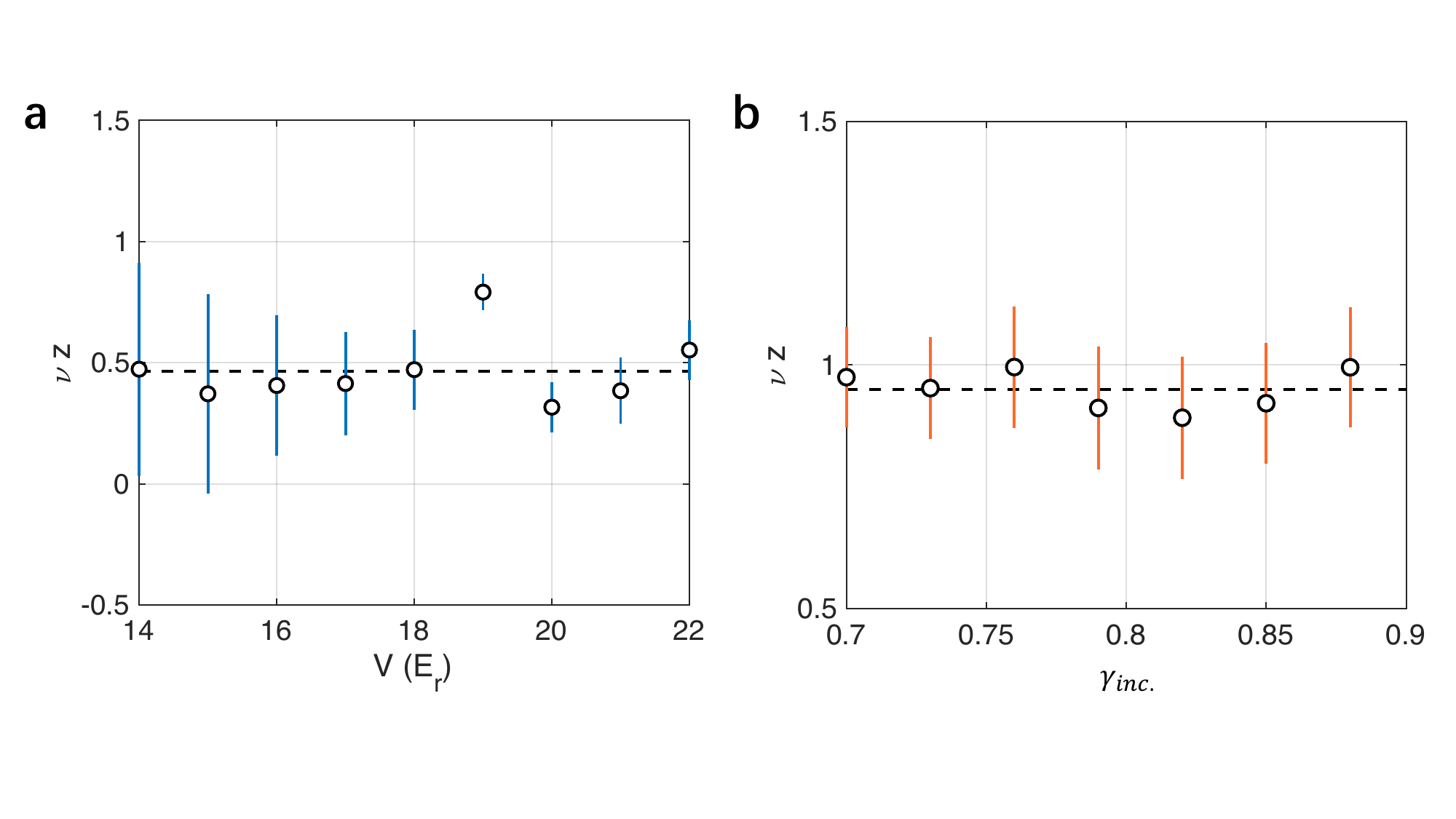}
	\caption{\textbf{Robustness of the {exponent} fit.} \textbf{a}, The quantum critical parameter $\nu z$ versus the cutting values of the trap depth $V$. $\nu z$ is obtained from the excitation fraction $n_{ex}(k)$. \textbf{b}, $\nu z$ versus the cutting values of $\gamma_{{inc}}$. $\nu z$ is obtained from the Mott insulators delay time $\tau_{MI}(k)$: the dashed lines show averaged value of data points on graphs. Error bars correspond to one standard deviation.}
	\label{SI_power}
\end{figure}

In the main text we show that by performing power law fitting to $\tau_{MI}$ versus $k$ and $n_{ex}$ versus $k$, different exponents are obtained. In principle, one could set an arbitrary cut bound to perform the fitting. Here we show that, among reasonable intervals, the cut bound does not {change} the exponents obtained. In Fig.~\ref{SI_power}\textbf{a}, we set different bound of potential depth, and the dependence of excitation fraction $n_{ex}$ on $k$ provides robust exponents and gives consistent $\nu z = 0.5 \pm 0.1$. And in Fig.~\ref{SI_power}\textbf{b}, in a range of $\gamma_{{inc}} \in [0.7,0.9]$, the delay time $\tau_{MI}$ versus $k$ data fits an exponent and gives consistent around $\nu z = 0.96 \pm 0.10$. Note that the $\nu z$ indicated here are averaged values of data points in Fig.~\ref{SI_power}, and are different from those in the main text, which are {analyzed} by cutting specific bounds. 

\subsection*{VII. GMFT simulation and the calculation of defect density}
We simulate the experiment with Gutzwiller Mean Field Theory (GMFT) \cite{Gutzwiller03, Kuba2005Gutzwiller}. The many-body wave function is written as a product state:
\begin{equation}
|\Psi \rangle = \prod_{i}\sum_{n}f^{n}_i |n\rangle.
\end{equation} 
The mean field Hamiltonian can be decoupled as products of single sites Hamiltonians:
\begin{equation}
\begin{aligned}
\hat{H}_{MF} & = \sum_{i,j}[-J(\psi_j a^{\dagger}_i + \psi^{\ast}_j a_i) + J\psi_i^{\ast}\psi_j] \\
&+ \frac{U}{2}\sum_{i}\hat{n}_i(\hat{n}_i-1) +  \sum_{i}  (\frac{1}{2}m\omega_0^2r_i^2-\mu) \hat{n}_i\\
& = \sum_i h_{mf},
\end{aligned}
\end{equation}
where $\psi_j = \langle \hat{a}_j\rangle$ refers to the local order parameter, $m$ denotes the particle mass and $\omega_0$ represents the vibrational frequency of the external harmonic trap. The simulation is performed in a 3D square lattice of $75^3$ site and all parameters are set to be exactly the same as those in our experiment. The chemical potential $\mu$ is set by the total particle number $N \sim 1.1 \times 10^5$. The local Hilbert space on every sites has cut-off $n_{max} = 3$ and the truncation errors are negligible. By solving minimization problem with respect to either $\langle \Psi |\hat{H} |\Psi \rangle$ or $\langle \Psi |\hat{H}-i\frac{\partial}{\partial t} |\Psi \rangle$, one could obtain the ground state under specific parameters or evolve the wavefunctions according to a time-dependent Hamiltonian respectively.

To resemble the {incoherent fraction} $\gamma_{{inc}}$ measured in the quasi-momentum profile, we define $\gamma_{{inc}}$ as the fraction except zero-momentum condensation:
\begin{equation}
\gamma_{{inc}} = 1 - \sum_{i,j}\frac{\langle a^{\dagger}_i\rangle \langle a_j\rangle}{N},
\label{MIfrac}
\end{equation}
where $N$ denotes total particle number. 

\begin{figure}[htbp]
	\centering
	\includegraphics[width=18cm]{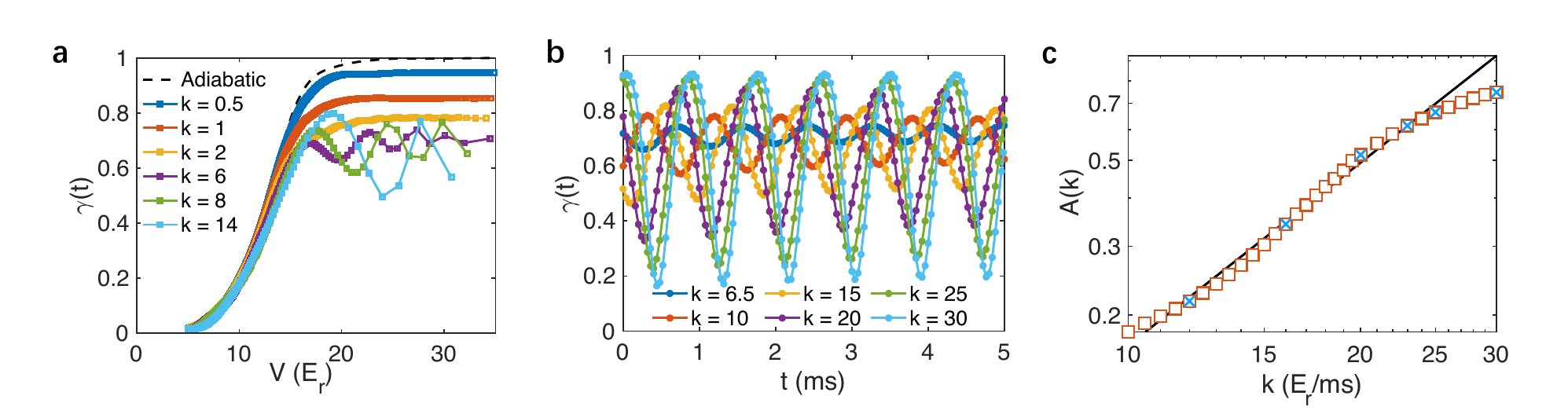}
	\caption{\textbf{GMFT results.} \textbf{a}, $\gamma_{{inc}}$ versus $V$ depending on different ramping rate $k$. For $k\le 4E_r$/ms, we see a retarded relaxation and $\gamma_{{inc}}$ starts to oscillate when $k>6E_r$/ms. The deviations of curves from the adiabatic one suggests that there are more coherent atoms remained on the Mott insulator background due to the external ramping. \textbf{b}, The oscillations of $\gamma_{{inc}}$ versus the holding time $t$. Following the same procedure in Fig.~4, we hold the sample at $V=25E_r$ for a time $t$. Different ramping rates $k$ result different oscillations of $\gamma_{{inc}}$. Based on this, we obtain the theoretical results in Fig.~4\textbf{c} in the main text. \textbf{c}, The oscillation amplitude at fast $k$ plot versus $k$. The linear fitting is performed in log-log space and an exponent of $1.56(6)$ is obtained for fitting all red squares on the plot. {In the main text Fig. 4c, we only list 6 points (blue cross in this panel c) with the same ramping rate $k$ in our data. If we only fit these 6 points (the same way as in Fig. 4), we will obtain an exponent at 1.41(15) which is mentioned in the main text.}}
	\label{SI_G}
\end{figure}

{The result of GMFT to compare with experimental data} is shown in Fig.~\ref{SI_G}, where the dynamics of different $\gamma_{{inc}}$ versus $V$ are obtained. We see the phenomena of a retarded relaxation and the phase oscillations in our GMFT calculations. However,
the {fail} of the thermalization {to unity} in the deep MI region is due to that the mean field theory fails to capture the strongly interacting systems, {and thus does not agree with experimental data}.
To calculate $\tau_{SF}$, we perform horizontal cut at $\gamma (t) = 0.6$ and find the coordinates of intersections $(V(k), 0.6)$ for each $k$, which is the same as in the data processing in our experiment. Excitation fractions $n_{ex}$ are extracted from $V = 18E_r$, same as the experimental data. For a fast ramp $k > 12 E_r$/ms, the amplitudes of the oscillations are calculated as the peak-to-peak values in Fig.~\ref{SI_G}\textbf{b}.

\begin{figure}[htbp]
	\centering
	\includegraphics[width=9 cm]{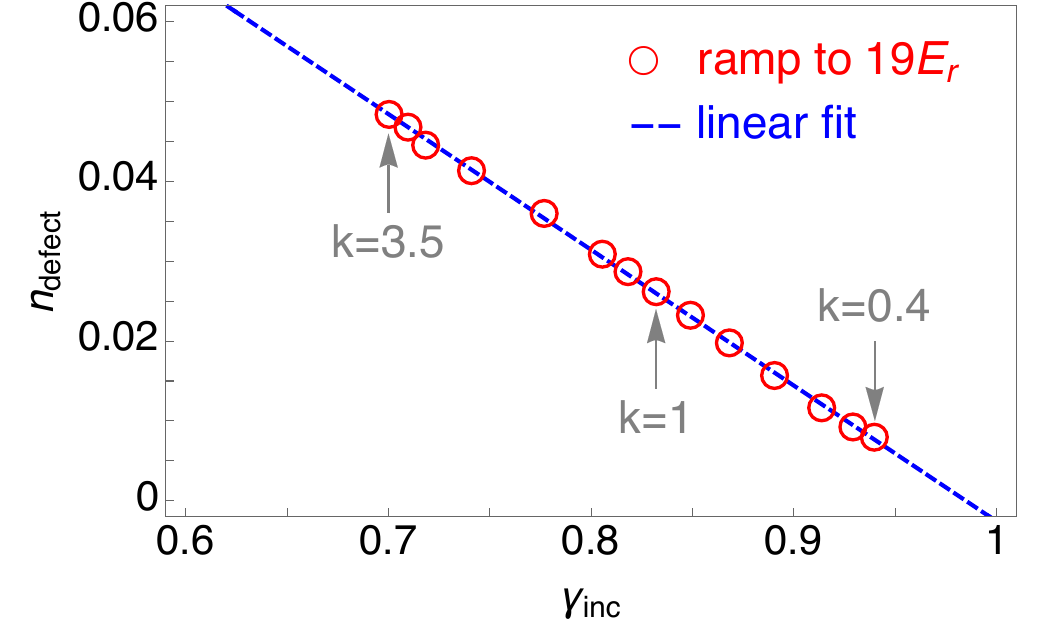}
	\caption{{\textbf{Defect density versus $\gamma_{inc}$ due to the external rampings.} 
	Each red circle corresponds to the ramping from $5E_r$ to $19E_r$ with one particular ramping rate $k$. $n_{\textrm{defect}}$ and $\gamma_{inc}$ are calculated by GMFT. Here we use gray arrows labeling the typical ramping rates $k=0.4$, 1, and 3.5$E_r$/ms. The blue dashed line is a linear fit with $n_{\textrm{defect}}=0.168(2)-0.170(2)\gamma_{inc}$.}
	\label{defect}}
\end{figure}

{The incoherent fraction $\gamma_{inc.}$ indicates the fraction of incoherent atoms.
Here we show theoretical calculations that $\gamma_{inc.}$ or $n_{ex}$ can be used to characterize the defect density in the system due to the the external ramping. We calculate the quantity of spatial defect density $n_{\textrm{defect}}=\langle (\hat{n}_i - \langle \hat{n}_i\rangle)^2\rangle$ by GMFT \cite{2012EDemler} for different ramping rate $k$. 
Here the inside bracket $\langle\hat{n}_i \rangle$ corresponds to the average of the particle number of the $i$-th site,
and the outside bracket $\langle$~$\cdot$~$\rangle$ corresponds to the average of the whole system.
Each time, we calculate $n_{\textrm{defect}}$ for the GMFT-evolved state at $V=19E_r$ for each particular $k$, and the numerical results are shown in Fig.~\ref{defect}. The ground state at $V=19E_r$ is a quantum-fluctuated Mott insulator whose $n_{\textrm{defect}}$ is not precisely zero, so the change of $n_{\textrm{defect}}$ versus $k$ is characterizing the additional excitations due to the external ramping. It shows a linear dependence on $\gamma_{inc}$, and this supports our measurement that $n_{ex}=\gamma_{inc}(adia)-\gamma_{inc}(k)$ is proportional to the excitations.}

{Figure 3 in the main text is not compared with GMFT simulations due to limit space in the main text figures. Here we add the simulation results in Fig. \ref{nex} to compare GMFT with our measurement results. The fitted exponents $\alpha$ from GMFT show consistency with our experimental observations in the regime of $13E_r$ to $19E_r$}
\begin{figure}[htbp]
	\centering
	\includegraphics[width=6 cm]{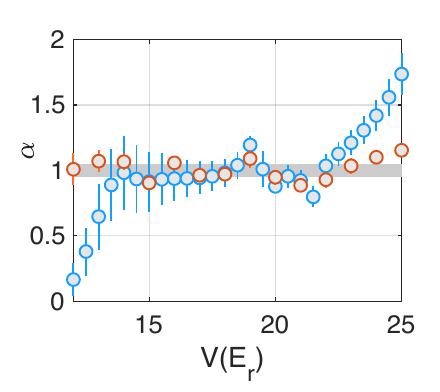}
	\caption{\textbf{GMFT simulations for Fig. 3b.} We add additional orange circles denoting GMFT results for the fitted exponent $\alpha$, while the experimental data are plotted by blue circles.}
	\label{nex}
\end{figure}

\end{document}